\documentclass[twocolumn,preprintnumbers,prb,superscriptaddress]{revtex4}
\def\bea{\begin{eqnarray}}
\def\eea{\end{eqnarray}}
\def\ben{\begin{equation}}
\def\een{\end{equation}}
\def\benu{\begin{enumerate}}
\def\enu{\end{enumerate}}

\def\sss{\scriptscriptstyle\rm}

\def\1var{(\bx_1...\bx\N)}

\def\ox{{\bf x}}

\def\br{{\bf r}}

\def\bx{{x}}

\def\x{_{\sss X}}
\def\c{_{\sss C}}
\def\s{_{\sss S}}
\def\xc{_{\sss XC}}
\def\Hxc{_{\sss HXC}}

\def\N{_{\sss N}}
\def\H{_{\sss H}}

\def\cresp{_{\rm c,resp}}
\def\resp{_{\rm resp}}
\def\ckin{_{\rm c,kin}}
\def\kin{_{\rm kin}}
\def\skin{_{\rm s,kin}}
\def\ext{_{\rm ext}}

\def\ee{_{\rm ee}}

\def\sph_int{ {\int d^3 r}}

\usepackage{graphicx}
\usepackage{psfrag}
\usepackage{amsmath}
\usepackage{bm}

\input epsf

\begin{document}
\title{Revisiting Molecular Dissociation in Density Functional Theory: A Simple Model}
\date{\today}
\author{David G. Tempel} 
\affiliation{Department of Physics and Astronomy, Hunter College and City University of New York, 695 Park Avenue, New York, NY 10065, USA\cite{1}}
\author{Todd J. Mart\'inez}
\affiliation{Department of Chemistry and The Beckman Institute, University of Illinois, Urbana, IL 61801, USA}
\author{Neepa T. Maitra}
\affiliation{Department of Physics and Astronomy, Hunter College and City University of New York, 695 Park Avenue, New York, NY 10065, USA}
\email{nmaitra@hunter.cuny.edu}

\begin{abstract}
A two-electron one-dimensional model of a
heteroatomic molecule composed of two open-shell atoms is considered.  
Including
only two electrons isolates and examines the effect that the highest
occupied molecular orbital has on the Kohn-Sham potential as the
molecule dissociates.  We reproduce the characteristic step and peak
that previous high-level wavefunction methods have shown to exist for
real molecules in the low-density internuclear region. The simplicity
of our model enables us to investigate in detail their development as
a function of bond-length, with little computational effort, and
derive properties of their features in the dissociation limit.  We
show that the onset of the step is coincident with the internuclear separation at
which an avoided crossing between the ground-state and lowest
charge-transfer excited state is approached. Although the step and peak features
have little effect on the ground-state energetics, we discuss their important
consequences for dynamics and response.
%We study several characteristic
%features arising in different components of the %exchange-correlation
%potential. These features include the ``step'' structure which
%develops in the response potential ($V_{resp}$) around the more
%eletronegative atom and the ``peak'' in the correlation kinetic
%potential ($V_{c,kin}$).  %Several analytic expressions for these
%features are also derived, %which become exact in the asymptotic limit
%of large inter-atomic separation.
%Lastly, we discuss the related
%problem of how approximate exchange-correlation functionals describe
%the dissociation process.

\end{abstract}
%\pacs{31.15.Ew,31.10.+z,31.25.Jf}
\maketitle

\section{Introduction}
The unprecedented balance between accuracy and efficiency of density
functional theory (DFT)~\cite{HK64,KS65,Kb99,PK03} is due in large
part to the discoveries of John Perdew. The mapping of the true system
of interacting electrons to a fictitious one in which the electrons
don't interact, yet reproduce the true electron density, requires
accurate approximations for the exchange-correlation (xc) potential, which
remained elusive until the developments in the 1980's of Perdew and
co-workers. Understanding and incorporating exact conditions and
physical principles underlie the robustness and reliability of Perdew's
functionals. In this spirit, we study here the structure of the exact xc potential as a molecule dissociates,
whose landscape of steps and peaks Perdew was one of the first to explore. 

%has become the standard approach to
%calculating ground-state electronic structure in a wide variety of
%systems, from atoms, molecules, and clusters, to nano-wires, and
%biomolecules~\cite{many}.
%In DFT a system of
%interacting electrons is mapped to a non-interacting Kohn-Sham (KS)
%system that is defined to yield the electron density of the true
%interacting electronic system. 
In DFT, one solves self-consistently the Kohn-Sham (KS) single-particle equations 
\ben 
(-\frac{1}{2}{\nabla^2} +
v\s[\rho](\br))\phi_i(\br)=\epsilon_i\phi_i(\br) 
\label{eq:kseqn}
\een 
where $v\s[\rho](\br)$ is the KS potential, a functional of the
ground-state electronic density, $\rho$. (Atomic units, $e^2 = \hbar = m_e =1$, are used
throughout this paper). It is usually written as the sum $v\s[\rho](\br) = v\ext[\rho](\br) + v\H[\rho](\br)
+ v\xc[\rho](\br)$, where $v\ext(\br)$ is the potential due
to the nuclei, $v\H(\br) = \int d^3r' \rho(\br')/\vert \br - \br'\vert$ 
is the classical Hartree potential and $v\xc(\br)$ is the
exchange-correlation (xc) potential, incorporating the remaining
many-body effects in a one-body potential. The KS orbitals
$\phi_{i}$ yield the true density according to \ben \rho(\br) =
\sum_{i=1}^{N}|\phi_{i}(\br)|^2 \label{eq:sum} \een In principle,
the ground-state density and all static properties of the true
interacting system are exactly recovered, but in practice approximations are
needed for the unknown xc potential $v_{xc}[\rho]$ as a functional of the density. 
% xc energy
%functional $E_{xc}[\rho]$, or equivalently its functional derivative,
Typically,
semi-local functionals, such as GGA's~\cite{PBE96} and meta-GGA's~\cite{SSTP03} give good energies and structural properties at equilibrium 
molecular
geometries; the non-empirical constructions of Perdew and co-workers impart a reliability to the description of diverse systems and properties. 
However, GGA's do not perform so well for weakly-coupled
sub-systems. Notably, recent work has been very successful in describing
van der Waal's forces using sophisticated non-local approximations in DFT~\cite{RDJS03,CTL08,VWV08}. For molecules dissociating into open-shells, the failure of
semi-local approximations becomes drastic, yielding unphysical
fractional charges on the separated species~\cite{Slater,PPLB82,P85b,RPCV06}.
This problem was first highlighted by
Perdew~\cite{PPLB82,P85b}, motivated by the observation of
Slater~\cite{Slater} that his ``$X\alpha$'' method yields a similar
result. 

Figure 7 of Ref.~\cite{P85b} shows that the exact xc potential
develops a ``region of positive constant'' around the atom with the
``tighter density distribution'', in the limit of infinite separation,
using a simple one-dimensional model.  
%This prevents the dissociation
%into fractionally charged fragments, and is closely related to the
%derivative discontinuity that the exact xc potential suffers as the
%number of electrons passes through an integer~\cite{PPLB82}. 
More generally, the effect of molecular dissociation on the
ground-state xc potential for the case of real diatomic closed shell
molecules consisting of open shell atoms has been studied
systematically, by Baerends, Gritsenko, and co-workers, in a series of
papers~\cite{BBS89,GLB96,GB96,GLB95,GLB94,GB97}. In Ref.~\cite{BBS89}, the
simplest case of this, the two-electron H$_2$ molecule was studied.
The absence of long-range left-right correlation in Hartree-Fock,
renders its potential overly repulsive near the nuclei, leading to an
overly diffuse density.  A highly accurate xc potential was
constructed from correlated CISD first- and second- order density
matrices in Ref.~\cite{BBS89}, and the resulting correlation potential
was shown to considerably reduce the repulsion at the nuclei.  It was
also shown that the xc potential develops a sharp maximum (``peak'')
at the bond midpoint.  A very thorough analysis of the KS potential in
stretched H$_2$ was performed later in Ref.~\cite{GB97}, where the
effect of different approximate constructions for the KS orbital was investigated and
explained in detail (see also Sec.~\ref{sec:asymptotic}). 
Using an iterative method introduced in
Ref.~\cite{LB94}, Ref.~\cite{GLB95} was the first to construct
molecular KS potentials for more than two electrons from correlated
densities. They studied LiH (and H$_2$) and found significant
differences with the local density approximation (LDA) at large
separations.  Ref.~\cite{GB96} calculated the xc potential for the
monohydrides XH (X=Li,B,F), analysing its structure via a
decomposition, or ``partitioning'' of $v\xc$ into various ``energy''
and ``response'' components related to the electronic
structure~\cite{BBS89,GLB94,GLB96}.  It was shown that left-right
correlation leads to a build-up in the xc potential around the H atom
(a ``step'', as was observed in the simple model of Ref.~\cite{P85b}).
 The ``peak'' present in the bond mid-point of H$_2$ was
found, in the case of the monohydrides, to shifted toward the H atom
while becoming significantly smaller due to the presence of core
electrons softening the left-right correlation effects.  The
partitioning scheme (reviewed in Sec.~\ref{sec:decomposition}), which had earlier been used to examine atomic xc
potentials~\cite{GLB94,GLB96}, proved to be a particularly useful tool
in the analysis, providing insight into the origin of the peak and
step structures.  
%We will review this partitioning briefly in Section~\ref{sec:decomposition}.

Molecular dissociation in DFT is particularly relevant when
considering time-dependent processes and nuclear dynamics on
potential energy surfaces.  The advent of time-dependent density
functional theory (TDDFT)~\cite{RG84,PGG96,TDDFTbook} allows for a
density-functional description of full electron dynamics and here
accurate long-range potentials are an important ingredient for many
applications, eg. photo-dissociation dynamics, excitations in large
molecules, including charge-transfer, and molecular
transport. A recent paper~\cite{LKQM06} discussed promising aspects
as well as challenges in getting accurate excited energy
surfaces from TDDFT;  certainly it is important to get the
ground-state potential energy surface correct.

% has received a renewed interest in
%recent years, with important implications for long-range excitations
%in Time-Dependent Density Functional Theory (TDDFT). In
%ref.~\cite{M05}, the `step-like' structure in the response potential
%($V_{resp}$) around the more electronegative atom of a dissociating
%molecule was found to lead to failure of the adiabatic approximation
%of TDDFT in describing long-range charge transfer excitations. In
%ref.~\cite{MT06}, the step was shown to lead to a failure of the
%adiabatic approximation for all long-range excitations,
%charge-transfer as well as local. Also of recent interest is the
%`peak' in the correlation kinetic potential $V_{c,kin}$ of a
%dissociating diatomic molecule, whose pressence has important
%implications for molecular transport (reference).

In the present paper, we study the xc potential of a dissociating
closed-shell hetero-atomic molecule consisting of two open-shell atoms
by analysing a simple one-dimensional model of two different
``one-electron atoms''.  The two ``electrons'' and two ``nuclei''
interact via soft-Coulomb interactions with the softening parameters
chosen to approximate certain properties of the real LiH molecule.
This simple model allows numerically exact
solution at a wide range of separations with little
computational effort, while reproducing the essential features, from the point of view of molecular dissociation, of the
xc potential for real three-dimensional molecules. It allows some analytic
treatment of these features that yields further insight into the
``step'' and ``peak'' structures mentioned above; for example, 
predicting the asymptotic height and position of the peak and an explanation 
of why such a structure, that hardly affects the energetics, must be there. 
A detailed examination of the stretched bond-length where the step
begins to appear, reveals a correlation with the position of the avoided
crossing between the ground-state and lowest charge-transfer states.
We explain why. 

A two-electron model isolates the effects due to the valence
electrons, which play the major role in dissociative processes,
without additional potential-features arising from core electrons.  In
the KS description of dissociation, the major role is played by the KS
HOMO, which, in the case of open-shell fragments, is delocalized
across the molecule.  By including only the HOMO in our model, we
isolate and examine effects on the dissociating potential energy
surface due solely to this most important orbital. The model is presented in Sec.~\ref{sec:model} while Sec.~\ref{sec:results} contain the numerical and analytic results.

We may draw conclusions from this simple model about real
three-dimensional molecules composed of open-shell fragments of
general odd electron-number, but with a little caution: we shall find
quantitative differences due to the lack of core electrons in our model, and 
the much
longer effective range of the soft-Coulomb interaction in 1D compared to the
true 3D Coulomb interaction. The soft-Coulomb interaction is used
in many interesting investigations of strong-field dynamics~\cite{JES88,EGLS92,VIC96,BN02,BL05,LGE00}, and recently, 
in the context of TDDFT~\cite{NB06,LK05}: these
models capture the essence of phenomena such as non-sequential
double-ionization, and laser-induced electron-recollision. 
%Our results here show that such models are based on
%KS potentials that have many quantitatively correct features carrying
%over for real molecules, but not all: for example the peak structure
%in these potentials is exaggerated compared to that found in real
%molecules with more than two electrons.  
The peak and step are challenging
features for approximations to capture, and are lacking in almost all
functionals used today. Being in a low-density
region, the peak structure has negligible energetic consequence,
however it does play a role when response or full dynamics is
considered: for example, it reduces the (hyper-)polarizability of
long-chain systems~\cite{KKP04}.  As they represent barriers to
electron transport, the work here is also relevant to one-dimensional
transport calculations in molecular wires~\cite{KCBC08}, although this fact has not been discussed before, so perhaps not yet been fully appreciated.  The step
structure, essential to avoid the fractional charge problem,  has severe consequences for the structure of the TDDFT xc kernel as we shall discuss in
Section~\ref{sec:implications}. 
%Our results complement those of Refs.~\cite{BBS89,GLB96,GB96,GLB95}.

\section{Decomposition of the xc potential}
\label{sec:decomposition} 
The partitioning of the xc potential~\cite{BBS89,GLB94,GB96,GLB95,GLB96} was motivated by first decomposing the xc energy components into ``potential'' and ``kinetic'' terms of the form:
\bea
\nonumber
E\xc[\rho] &=& W\xc[\rho] + T\c[\rho]\, \;\;\;\; {\rm where}\\
\nonumber
W\xc[\rho] &=& \frac{1}{2}\int \rho(\br) v\xc^{\rm hole}[\rho](\br) d^3r \;\;\;\; {\rm and}\\
T\c[\rho]  &=& \int\rho(\br) v\c^{\rm kin}[\rho](\br)
\eea
implicitly define ``hole'' and ``correlation kinetic'' potentials, $v\xc^{\rm hole}$ and $ v\c^{\rm kin}$. The total xc potential is then partitioned into 
 three components:
\ben
v\xc[\rho](\br) = v\xc^{\rm hole}[\rho](\br)+v\ckin[\rho](\br)+v\resp[\rho](\br)
\label{eq:partition}
\een
$v\xc^{\rm hole}[\rho](\br)$ is the Coulomb potential of the xc
hole:
\ben
v\xc^{\rm hole}(\br) = \int \frac{\rho\xc(\br, \br_{2})}{|\br - \br_{2}|}d^3\br_{2}
\label{eq:vxchole1}
\een
where the xc hole $\rho\xc(\br_{1}, \br_{2})$ is defined through the pair density $P(\br_1,\br_2)$ (joint probability of finding an electron at $\br_1$ while another is at $\br_2$), via $P(\br_1,\br_2) = \rho(\br_1)(\rho(\br_2) + \rho\xc(\br_{1}, \br_{2}))$. 
When added to the Hartree potential, $v\xc^{\rm hole}(\br) + v\H(\br)$
represents the average repulsion an electron at position $\br$ experiences
due to the other (N-1) electrons in the system.
%seen by introducing the conditional probability amplitude $\Phi(s_1,
%x_2,...,x_N|r_1)$. For an electron at position $r_1$, in terms of the
%full many-body wavefunction $\Psi(x_1,x_2,...,x_N)$
In terms of the conditional probability amplitude, whose square gives the probability of
finding the other (N-1) electrons in the system with space-spin
coordinates $\ox_2,\ox_3,...,\ox_N$ when an electron is known to be at
position $\br_1$:
\ben
\Phi(s_1, \ox_2,...,\ox_N|\br_1) = \frac{\Psi(\ox_1,\ox_2,...,\ox_N)}{\sqrt{\frac{\rho(\br_1)}{N}}}
\label{eq:condit}
\een
where $\Psi(\ox_1,\ox_2,...,\ox_N)$ is the interacting many-electron wavefunction,
we have
\begin{eqnarray}
\nonumber
&&v\xc^{\rm hole}(\br)+ v\H(\br)= \int \Phi^*(s_1, \ox_2,...,\ox_N|\br) \\  &&\times \left [\sum_{i=2}^{N}\frac{1}{|\br-\br_i|}\right ] \Phi(s, \ox_2,...,\ox_N|\br)ds_1d\ox_2...d\ox_N
\label{eq:vxchole}
\end{eqnarray}

The second term in Eq.~(\ref{eq:partition}), $v\ckin[\rho](\br)$, is
the correlation contribution to the kinetic component of the xc
potential. It is the difference of the kinetic components of the
interacting and non-interacting KS systems: \ben v\ckin(\br) =
v\kin(\br)-v\skin(\br) \label{eq:vckin} \een where the kinetic
components may be written in terms of the conditional probability
amplitude: 
\bea 
\nonumber v\kin(\br_1) &=&
\frac{1}{2}\int {\left \vert \nabla_1 \Phi(s_1,
\ox_2,...,\ox_N|\br_1) \right \vert}^2 ds_1d\ox_2...d\ox_N\\ &=&
\frac{\nabla_{1'}\nabla_{1}\rho(\br_{1}', \br_{1})|_{{\br_{1} =
\br_{1}'}}}{\rho(\br_{1})} - \frac{[\nabla \rho(\br_1)]^2}{8
\rho(\br_{1})^2} \label{eq:vkin} \eea 
and 
\bea \nonumber
v\skin(\br_1) &=& \frac{1}{2} \int {\left \vert \nabla_1 \Phi_s(s_1, \ox_2,...,\ox_N|\br_1) \right \vert }^2 ds_1d\ox_2...d\ox_N\\
&=&\frac{1}{2}\sum_{i=1}^{N}\left\vert\nabla_{1}\frac{\phi_{i}(\br_{1})}{\rho^{\frac{1}{2}}(\br_{1})}\right\vert^2
\label{eq:vskin}
\eea
In Eq.~(\ref{eq:vkin}) $\rho(\br_1', \br_1)$ is the first-order spin-summed reduced
density-matrix, and in Eq.~(\ref{eq:vskin}) $\Phi_s(s_1, \ox_2,...,\ox_N|\br_1)$ is the conditional probability
amplitude of the KS system, which is defined as in Eq.~\ref{eq:condit} but
with the
KS single Slater determinant, whose orbitals are $\phi_i(\br)$, replacing the full many-electron wavefunction.
%$\Psi_s(x_1, x_2,...,x_N|r_1)$
%according to:
%
%\ben
%\Phi_s(s_1, x_2,...,x_N|r_1) = \frac{\Psi_s(x_1,x_2,...,x_N)}{\sqrt{\frac{\rho(r_1)}{N}}}
%\een

The final term in Eq.~(\ref{eq:partition}) is the so-called response
potential. It may be further partitioned into terms representing the response
of the xc hole, and the response of the correlation kinetic
potential ~\cite{GLB96,GB96}, but we will not pursue this further decomposition here. 

For two electrons in a spin-singlet, many of these expressions
simplify considerably, as the KS single-Slater determinant consists of
one doubly-occupied spatial orbital, $\phi_0(\br)$, which is equal to
square root of half the density:
\ben
\phi_0(\br)=\sqrt{\frac{\rho(\br)}{2}}
\label{eq:orbital}
\een
Substituting into Eq.~\ref{eq:kseqn}, we can solve explicitly for the KS potential as a functional of the density:
\ben
v\s[\rho](\br)=\frac{\nabla^2\sqrt{\rho(\br)}}{2\sqrt{\rho(\br)}}-I
\label{eq:kohn-sham}
\een
where $I$ is the first ionization potential of the system.
From Eq.~(\ref{eq:vskin}), $v\skin(\br)=0$ and Eq.~\ref{eq:vckin} reduces to:
\ben
v\ckin(\br) = v\kin(\br)
\een
Also, for two electrons,
$v\x(\br) = {v\x}^{\rm hole}(\br) = -\frac{1}{2}v\H(\br)$.
The exchange component to the response potential, $v\resp$ is zero. We may therefore write:
\bea
\label{eq:vx}
v\x(\br)&=& -v\H(\br)/2 =  v\x^{\rm hole}(\br)\\
v\c(\br)&=& v\c^{\rm hole}(\br) + v\ckin(\br) + v\cresp(\br)
\eea
%i.e. the exchange
%contribution to the xc hole in our analysis is simply the self-interaction correction potential $-\frac{1}{2}V_H(r)$

%%David, we don't calculate these components separately, right? Then will cut or shorten this next part, just referring to lit.
%\ben
%V_{resp}[\rho](r) = V_{xc}^{hole,resp}[\rho](r) +V_{c,Kin}^{resp}[\rho](r)\;,
%\een
%where
%\ben
%v\xc^{\rm hole,resp}[\rho]({r_{3}})= \frac{1}{2}\int \frac{\rho(r_2) \rho(r_1)}{|r_{1} - r_{2}|} \frac{\delta g([\rho]; r_{1}, r_{2})}{\delta \rho(r_{3})}d^3r_{1} d^3r_{2}
%\een
%and
%\ben
%V_{c,kin}^{resp}[\rho](r_{1}) = \int \rho(r_{2}) \\ x \frac{\delta}{\delta \rho(r_{1})} \left [V_{kin}[\rho](r_{2}) - V_{s,kin}[\rho](r_{2})\right ] d^3r_{2}
%\een

As was found in Refs.~\cite{GLB96,GB96}, as a heteroatomic molecule
dissociates, a step structure in the low-density bond midpoint region
arises in the response component $v\resp$, accompanied by a peak
structure in the kinetic component $v\ckin$ of the xc potential.  (See
also Figures~\ref{fig:vxcholekinresp} and \ref{fig:R10pots}).  

Consider now a
simplified description of the molecule that has includes just one
electron on each atom.  As explained in Ref.~\cite{BBS89}, and
reflected in Eq.~\ref{eq:vkin}, $v\kin(\br)$ depends on the gradient of the conditional probability amplitude, so describes how strongly
the motion of an electron at reference position $\br$ is correlated
with the other electrons in the system.  For $\br$ near one of the
nuclei, the reference electron moves in a potential dominated by the
nuclear potential, and the conditional amplitude $\Phi$ reduces to the atomic HOMO of the other atom, and doesn't change for small changes around $\br$; hence $v\kin$ goes to zero. But in the
internuclear region, the motion of the two electrons becomes
correlated: as the reference position moves from one nucleus to the
other, the conditional probability of finding the other switches from
being towards one atom to the other, and so $v\kin$ peaks.

 The origin of the step structure was also analyzed extensively
 in~\cite{GB96} and shown to arise in the correlation component of the
 response potential, $v\c^{\rm resp}$. It was discovered
 earlier~\cite{P85b,AB85}, in relation to the derivative
 discontinuity~\cite{PPLB82,P85b,AB85,PL97}, that the correlation
 potential for a long-range molecule composed of two open-shell atoms
 must have a step in the midpoint region, whose size is such that the
 atomic HOMO orbital energies re-align.  From Koopman's theorem, the
 HOMO energy equals the ionization potential; therefore the step has a
 size $\Delta I = I_2 - I_1$ where $I_{2,1}$ is the larger(smaller)
 ionization potential of the two atoms, raising the potential of the
 more tightly bound atom.  Far away from the molecule the potential
 near this atom steps back down to zero.  A simple way to understand
 the origin of the step is to realise that had the step not been
 there, then one could lower the ground-state energy of the long-range
 molecule by transferring a fraction of charge from the atom with the
 higher ionization potential to that with the lower, leading to the
 molecule dissociating into fractionally charged species. As this
 cannot happen, the KS potential develops a step in the bonding
 region, which re-aligns the atomic HOMO's, so preventing any
 bias. Another way to put this, is that the chemical potential must be
 the same throughout the long-range molecule, and equal to the
 molecular HOMO orbital energy. Since the chemical potential of the
 true system is the smallest ionization potential in the system at infinite separation, the
 KS potential near the atom with the larger atomic ionization
 potential must be uniformly raised by $\Delta I$ to bring it to the
 ionization potential of the other atom, while asymptotically stepping back down
to zero. 

%(*tidy*) As the step occurs in a region of very low density, it itself 
% doesn't affect the energetics of the ground-state -- however, lack of the step n local/semi-local approximations leads to distorted atomic potentials that contain a fractional number of electrons on the separated species. 

%The effect of the step and peak on dynamics and excited state properties is discussed in Sec.~\ref{sec:implications}.

 We shall now introduce our two-electron model to study these features
 further and how they develop as a function of bond-length.

\section{A One-Dimensional Two-Electron Model of LiH}
\label{sec:model}
A simple one-dimensional, two-electron model of lithium hydride can
be used to illustrate several important features of hetero-atomic
dissociation. Much of the essential physics of the dissociation
process may be captured by focusing on the chemically important
valence electrons, while representing the effect of the core
electrons by an average effective potential, such as a
pseudopotential or frozen-core approximation. In the case of LiH,
the two core electrons are localized in the Li 1s shell, while the
two valence electrons are delocalized across the molecule. Our
goal is to analyze the effect of bond breaking and formation on the
various xc components (Eq.4), which in a  real molecule will
be partially obscured by shell structure and other many-electron effects from 
the electrons in the Li 1s core. Our two-electron model enables us
to circumvent this complication, by focusing solely on the electrons
involved in the bond, and their effect on the Kohn-Sham
characteristics. As further simplication, it is reasonable to use a one-dimensional model,
where the coordinate is taken to be along the bond axis, for cylindrically symmetric systems such as a
diatomic molecule.

As is often done in one-dimensional models, the Coulomb potential
$\pm 1/\vert \br - \br'\vert$ is replaced by a soft-Coulomb
potential, $\pm 1/\sqrt{a + (x-x')^2}$. For a model of LiH at
interatomic separation $R$, we write the electron-nuclear potential as: 
\ben
v\ext(x) =-\frac{1}{\sqrt{a+(x-R/2)^2}}-\frac{1}{\sqrt{b+(x+R/2)^2}}
\label{eq:vext} 
\een 
The ``softening parameters'' $a$ and $b$ are
directly related to the ionization potentials of the individual
atoms (see shortly). Similarly, the electron-electron repulsion is represented by
a soft-Coulomb form: \ben v\ee(x) =
\frac{1}{\sqrt{c+(x_{1}-x_{2})^2}} \label{eq:vee} \een

\begin{figure}[h]
 \centering
 \includegraphics[height=4.5cm,width=6cm]{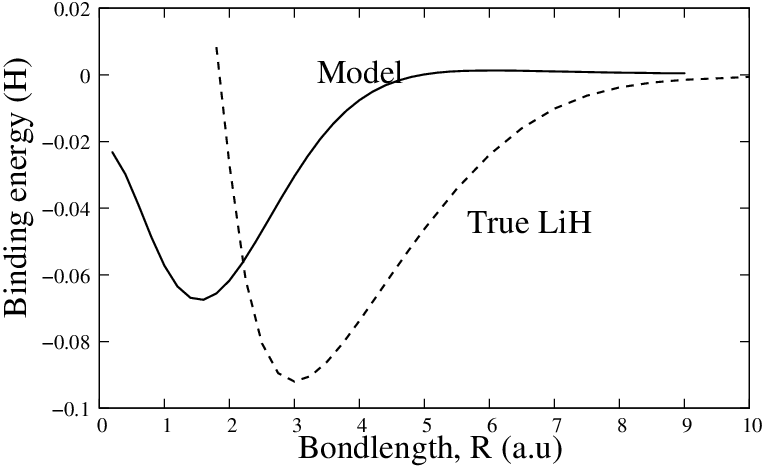}
\caption{Binding energy for: 1) model 1D LiH 2) true 3D LiH}
\label{fig:disscurve}
\end{figure}

We place Li at $-R/2$ and H at $R/2$, and choose the parameters
$a=0.7, b=2.25$ and $c=0.6$, for reasons explained in the following.
With $a=0.7$, the
%Li is placed at coordinate $-R/2$ with
%a softening parameter of 2.25 while H is at $+R/2$ with softening
%parameter .7.
ionization potential of hydrogen in our model comes out to be
$0.776$H as compared with $0.5$H for the real atom. Taking $b=2.25$
yields that of Li in our model as $0.476$H as compared with $0.198$H
for the real lithium atom. The correct difference in ionization
potentials of the atoms $\Delta I = I_{H}-I_{Li} = 0.3$H is however
exactly reproduced by our parameters; $\Delta I$ is
a key quantity in our analysis of the KS potential at large interatomic separations.
%structure  in $V_{resp}$,
%the peak in $v\ckin$ and also the structure of the xc hole (**).
Due to the long-range nature of the soft-Coulomb interaction, we choose 
the atomic ionization potentials be larger than in
the true 3D case to prevent the atomic densities of the individual atoms
from being too diffuse. Other factors considered were the
equilibrium bond length (model $1.6$a.u., true 3.0a.u), dissociation energy (model
$0.068$ H, true 0.092H) and molecular first ionization potential (model $0.51$H, true 0.29H), where the nuclear-nuclear interaction is modelled by
\ben
v_{nn}(R) = \frac{1}{\sqrt{(a+b-c) +R^2}}
\een 
In
Fig.(\ref{fig:disscurve}) the dissociation curve for our model is
plotted for comparison with that of 3D LiH~\cite{BG92b}.

%\begin{figure}[h]
% \centering
% \includegraphics[height=5cm,width=6cm]{curve.eps}
%\caption{Total groundstate energy for: 1) 1-d Soft Coulomb LiH 2) 3-D LiH}
%\label{fig:curve}
%\end{figure}

\section{Numerical Solution and Results}
\label{sec:results}
%From the Hohenberg-Kohn theorem~\cite{HK64,KS65}, the KS potential
%is a unique functional of the
%density, but in general the explicit form of this functional
%dependence is unknown. But for the special case of two electrons
%in a spin-singlet, there is only one doubly-occupied KS
%orbital, Eq.~(\ref{eq:orbital}), which upon substitution into Eq.~(\ref{kohn}), directly yields
% As this orbital to be real, we can solve
%eq.(~\ref{eq:sum}) explicitly, which yields the expression in
%eq.(~\ref{eq:orbital}). Then Substituting eq.(~\ref{eq:orbital}) into
%the KS equation eq.(~\ref{kohn}) and solving for the KS
%potential, one obtains:

We use a standard Runge-Kutta differential equation solver, as
implemented in the octopus code~\cite{octopus,octopus2,octopuswebsite}
%David: Also explain numerical convergence issues at large separation, difficulties with going too far and why
to numerically solve for the ground-state wavefunction $\Psi(x,x')$
of the Hamiltonian: \ben H=
-\frac{1}{2}\frac{d^2}{d{x_1}^2} - \frac{1}{2}\frac{d^2}{d{x_2}^2} +
v_{ext}(x_1) + v_{ext}(x_2) + v_{ee}(x_1-x_2) \een where $v\ext(x)$
and $v\ee(x_1-x_2)$ are defined in Eqs.~(\ref{eq:vext}) and
(\ref{eq:vee}). The above two particle Hamiltonian is mathematically
equivalent to that of one particle moving in the two dimensional
potential~\cite{octopus,octopus2,octopuswebsite}: \ben v\ext(x) + v\ext(y) + v\ee(x-y) \een
 We solve the equivalent one-particle Schr\"odinger equation on a
rectangular two dimensional 25 by 25 a.u. real space grid. The grid
points are separated by a distance of .04 a.u.
%Because the
%schrodinger equation in our model is not computationally demanding,
%with impunity we chose the grid size to be much larger than the
%dimensions of the system under consideration to prevent numerical
%problems from ``edge effects''. Similarly, the grid spacing is
%chosen to be sufficiently small to ensure proper convergence without
%concern for computational demand.
%Due to numerical error arising in the region of exponentially small
%overlap between the atoms, it is not possible to obtain meaningful
%solutions beyond an inter-atomic separation of 9 a.u. However,
%because we are concerned with 
%The features of the xc potential
%asymptotic limit for xc properties is already reached at a
%separation of 10 a.u. which will be considered ``asymptotic'' in the
%remainder of our analysis.

The density is obtained from the wavefunction through $\rho(x) =
2\int dx' \vert\Psi(x,x')\vert^2$, and then substituted into
Eq.~(\ref{eq:kohn-sham}) to yield the exact KS potential.
The xc potential can be isolated from subtracting the
external potential Eq.~(\ref{eq:vext}) and the Hartree potential, $v\H(x) = \int \rho(x)v\ee(x-x')dx'$ (using $v\ee$ from Eq.~(\ref{eq:vee})). Because $v\x(x) = -v\H(x)/2$, we may also extract the
correlation potential alone $v\c(x)$.  
From the conditional probability amplitude~\ref{eq:condit}, we construct $v\xc^{\rm hole}(x)$ according to Eq.~(\ref{eq:vxchole}).

\begin{figure}[h]
 \centering
 \includegraphics[height=8.5cm,width=8cm]{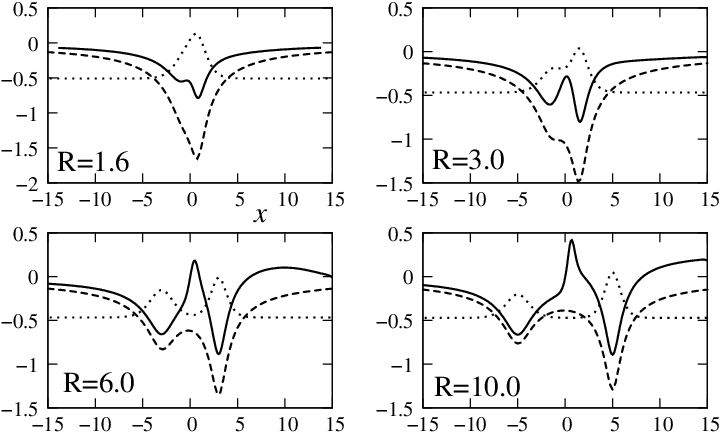}
\caption{$v\s$ (solid curves), $v\ext$, and the density (dotted curves) plotted at the internuclear separations indicated.}
%a) 1.6 a.u. (equilibrium bond-length) b) 3.0 a.u. c) 6.0 a.u. d) 10.0 a.u.}
\label{fig:vsvextdens}
\end{figure}

The exact KS potential is plotted at several different internuclear
distances in Fig.(\ref{fig:vsvextdens}) alongside the external potential
and the density. As the molecule dissociates, step-like and peak-like
features clearly develop in the KS potential. There is a build-up in
the KS potential around the more electronegative atom that, at each $R$, eventually returns to zero
on the right-hand side of the atom (one sees the beginning of the return to zero at the smaller separations shown, but at separation $R=10.0$ this occurs beyond the region plotted).

\begin{figure}[h]
 \centering
 \includegraphics[height=8.5cm,width=8cm]{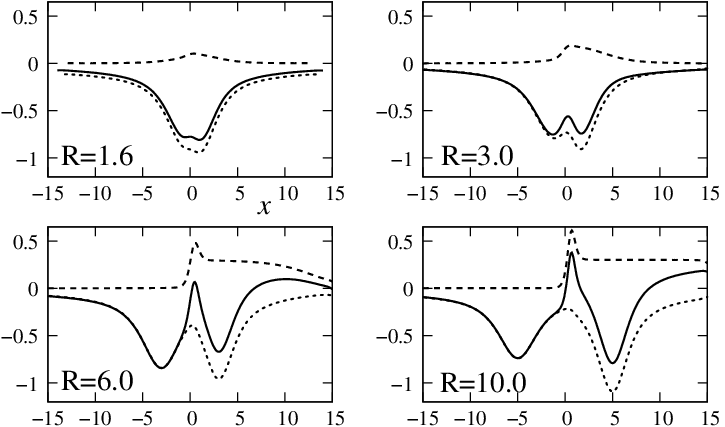}
\caption{The total xc potential $v\xc$ (solid curves), $v\ckin+v\resp$ (dashed curves),$v\xc^{\rm hole}$ (dotted curves) at various internuclear separations}
\label{fig:vxcholekinresp}
\end{figure}

These features occur in the response and kinetic components of the
correlation potential, as is evident in Fig.~\ref{fig:vxcholekinresp}. Here,
we plot the xc potential (solid), which is the sum of the xc-hole
potential (dotted) and the response components $v\ckin+v\resp$
(dashed).
At equilibrium bond-length ($R=1.6$a.u.) the xc potential is
dominated by the potential of the xc hole. As the molecule
dissociates the $v\ckin + v\resp$ components become large giving
rise to clear peak and step structures. At large separation, the local
variation of the total xc potential around each atom is almost entirely
due to the xc hole: at $R=6.0$a.u. and 10.0 a.u., $v\xc$ and
$v\xc^{\rm hole}$ exactly coincide near Li, while near H they have the
same shape, but the well in $v\xc$ is translated upward by exactly
3.0 a.u. relative to $v\xc^{\rm hole}$, which is the magnitude of the step $\Delta I$ (see earlier Sec.~\ref{sec:decomposition}). 
At $R=6.0$a.u., 
the step has reached its asymptotic value of $I_H -I_{Li} = 3.0$a.u.
%This can be seen most clearly in the dashed curve
%of fig.(\ref{fig:3comp}), where the sum of $V_{c,kin}$ and $V_{resp}$
%is plotted.
As the molecule is pulled apart further, the step does not increase in
size, but becomes flatter and larger in spatial extent.  In
Section~\ref{sec:pes}, we show that the bond-length at which the step
begins to develop is related to the position of the avoided crossing
between the ground-state and the state that eventually becomes the lowest charge-transfer state. 

%At equilibrium bondlength ($R=2.6$a.u.), $V_{xc}^{hole}$ is slightly
%deeper near Li. As the molecule dissociates, there is a ``switching
%over'', so that at larger separations $V_{xc}^{hole}$ is significantly
%deeper on H.
%David - i am not sure what the relevance is of this deep-ness. I'd have thought the effective area would be more relevant. Probably just cut the above unless discussed for many-electron true case in previous literature - check to see.

A sharp peak near the rise of the step is evident in the xc potential
(Fig.~\ref{fig:vxcholekinresp}); this occurs in the kinetic component
to the correlation potential, $v\ckin$, as discussed in Sec.~\ref{sec:decomposition}.
 We return to an analysis of its magnitude and location in the widely
separated limit, and an explanation of its role in achieving the exact density of the interacting system, in Sec.~\ref{sec:asymptotic}.

%%David -can you also plot the difference of Vc at the nuclear positions as a function of R; also do the same for v_{c,kin} + v_{resp}

\begin{figure}[h]
\centering
 \includegraphics[height=5cm,width=7cm]{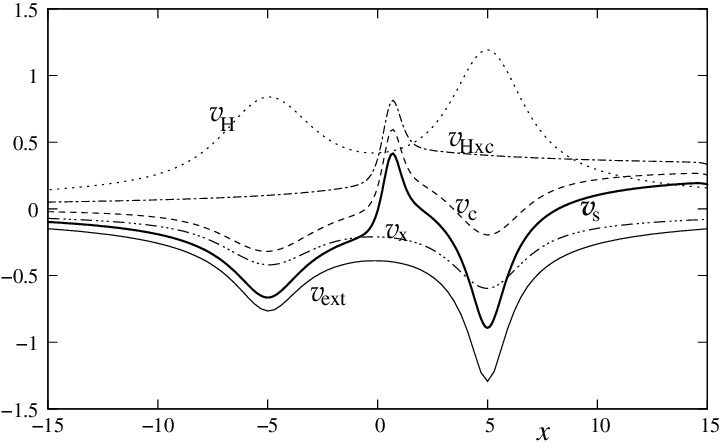}
\caption{Components of the potentials for  $R=10.0$}
\label{fig:R10pots}
\end{figure}

In Figure~\ref{fig:R10pots}, we plot the potentials for the separation
$R=10.0$.  This is the largest separation for which we could converge
our numerical method. In the limit of very large separation, we expect
that the KS potential reduces simply to the external potential in the region
of the nuclei, because it would
be a one-electron system around each nucleus. There may be a possible shift up or down relative to the external potential, since constants in potentials have no physical relevance. 
 That is, we expect the Hartree-plus-xc
potential becomes flat in the atomic regions.  We notice in our model
at R=10.0, that this is approximately true: there is however a gentle
slope in $v\Hxc$, upward around the left nucleus, and downward on the right, and
this is largely due to a Hartree effect.
Compared to the atomic densities in true
Coulomb-interacting systems, the soft-Coulomb densities in
one-dimension fall off much slower away from their nuclei, resulting
in a longer-ranged Hartree and xc potential than in the true 3D counterpart.
It is clear from the graph that the Hartree potential is
still significant in the interatomic region. In addition to long-ranged
correlation effects from the density on the ``other'' atom (i.e. the peak and step), the xc
%exchange potential (which is half the Hartree) and correlation 
potential must cancel the local Hartree potential: the exchange potential takes care of half of this cancellation (Eq.~\ref{eq:vx}), but the correlation potential must
also contribute a well of half the size of the Hartree, as is evident in the graph. Despite the long-rangedness of the Hartree potential, $R=10.0$ can still be viewed as ``asymptotic'' from the point of view of the peak and step structures in the correlation potential: the graph shows clearly that the potential on the hydrogen nucleus on the right is raised by $\Delta I$, and the peak has a height of about 0.76 (see last section).

\subsection{Onset of the step: Relation to Potential Energy Surface Crossings}
\label{sec:pes}
We now show that the bond-length at which the step begins to
significantly develop is correlated with the position of the avoided
crossing in the potential energy surfaces associated with the ground
state and the lowest excited charge transfer state. In a diabatic
picture, ionic and covalent curves cross at an internuclear distance,
$R\c$, which is approximately equal to $1/(I_D - A_A)$, where $I_D$ is the
ionization energy of the donor and $A_A$ the electron affinity of the
acceptor, in the lowest charge-transfer state of the long-range
molecule~\cite{GH74}. When one considers the adiabatic potential
energy surfaces, the crossing becomes an avoided one, whose splitting
exponentially decreases as a function of $R\c$~\cite{GH74}.

What has not been previously pointed out, however, is that the step
structure in the KS potential begins to develop in the vicinity of the
avoided crossing.  Why this must be so lies in the fact that the step
is an asymptotic feature, that arises once the two atoms are
independent systems, and its shift of the eigenvalues of the more
tightly bound atom ensures that the ground state solution of the KS
potential has exactly half the density (i.e. one electron) on either
side of the midpoint (see Sec.~\ref{sec:decomposition}). The
development of the step must therefore track the independence of the
two atomic systems (measured, for example, by their indifference to a
perturbation on the other atom). The avoided crossing marks the point
at which the molecule transitions (moving from short bond distances to
longer ones) from a single system to two independent systems.  The
width of this transition tracks the magnitude of the ground-excited
energy gap at the avoided crossing, i.e. it should be wider when the
avoided crossing is at small bond distances and sharper when the
avoided crossing occurs at large distances.

\begin{figure}[h]
 \centering
 \includegraphics[height=8cm,width=7cm]{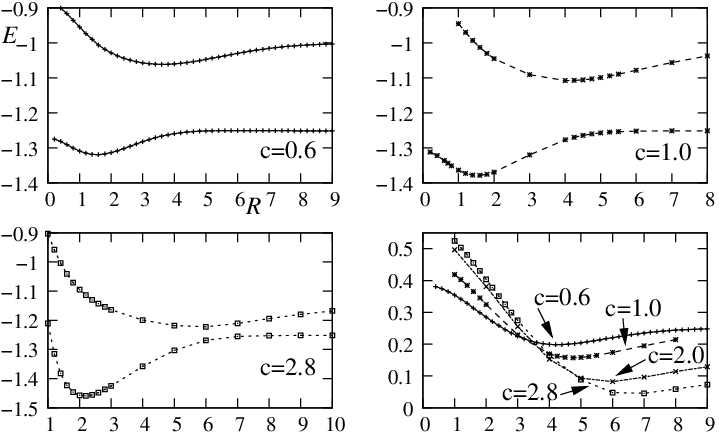}
\caption{Ground-state and first-excited state (charge-transfer) potential energy surfaces for our model with $c=0.6$ (top left),$c=1.0$ (top right),$c=2.8$ (bottom left). The energy differences between the surfaces are shown in the bottom right figure; their minimum lies at the avoided crossing.}
\label{fig:pes}
\end{figure}

Our model demonstrates this explicitly. 
Figure~\ref{fig:pes} presents the ground- and first excited-state
potential energy surfaces for three different values
of the electron-electron soft-Coulomb parameter, $c$. As $c$
increases, the avoided crossing moves out and becomes  sharper; the lowest energy gap therefore decreases, indicating that the transition from ionic to covalent character occurs more abruptly. 

\begin{figure}[h]
 \centering
 \includegraphics[height=5cm,width=7cm]{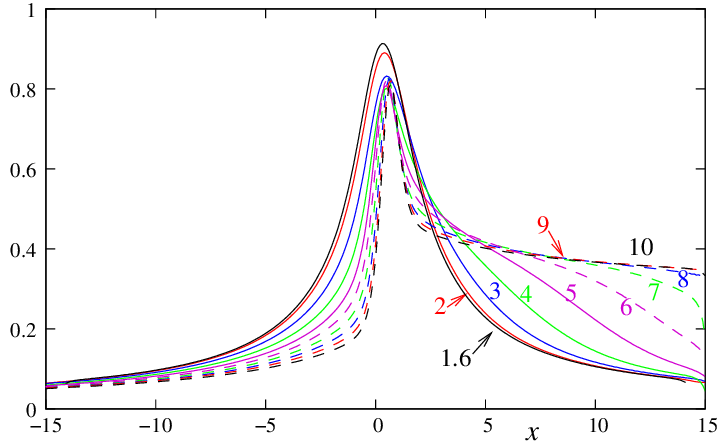}
\caption{The Hartree-exchange-correlation potential, $v\Hxc(x)$ for our LiH model ($c=0.6$); the values of interatomic separation $R$ are indicated. }
\label{fig:Vhxcstepa}
\end{figure}

In Figure~\ref{fig:Vhxcstepa}, we plot the
Hartree plus xc potential, $v\Hxc(x) = v\H(x) + v\xc(x)$  for a range of internuclear
separations $R$, for $c=0.6$. As this is the net potential that gets
added to the external potential, we expect that in the limit of wide
separation, it becomes flat around each nucleus, since it should
describe essentially two one-electron systems. We see this in the graph, where a definite step is visible from $R=5.0$ and higher.
We see that it is indeed in the approach to the avoided crossing, at about R=4.0, that a shoulder first becomes clearly visible around the atom with the higher IP; this develops fully into a step of size $\Delta I$, as the molecule dissociates.  

\begin{figure}[h]
 \centering
 \includegraphics[height=5cm,width=7cm]{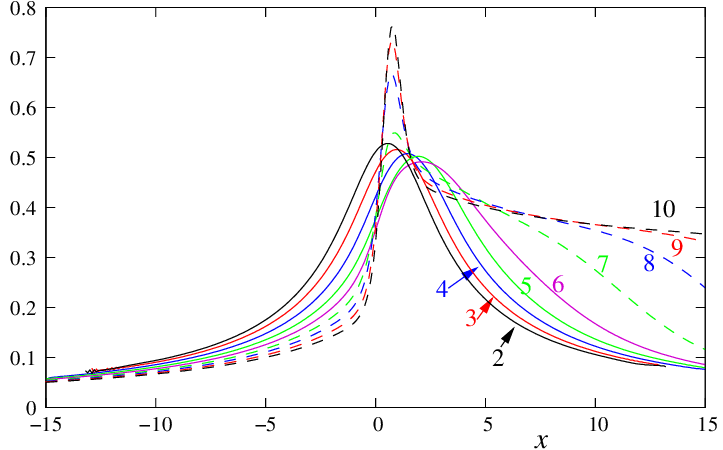}
\caption{As in Fig. ~\ref{fig:Vhxcstepa} but with $c=2.8$}
\label{fig:Vhxcstepc2.8}
\end{figure}

In Figure~\ref{fig:Vhxcstepc2.8}, we plot $v\Hxc$ for $c=2.8$. The step begins to
develop at larger $R$, corresponding to the larger $R\c$ where the
avoided crossing occurs. Also, as the avoided crossing becomes sharper, 
the onset of the step happens more rapidly. 

%\begin{figure}[h]
% \centering
% \includegraphics[height=6cm,width=7cm]{Figures/Vhxcstepe.eps}
%\caption{The Hartree-exchange-correlation potential for our model but with $c=1.0$}
%\label{fig:Vhxcstepe}
%\end{figure}

\begin{figure}[h]
 \centering
 \includegraphics[height=5cm,width=7cm]{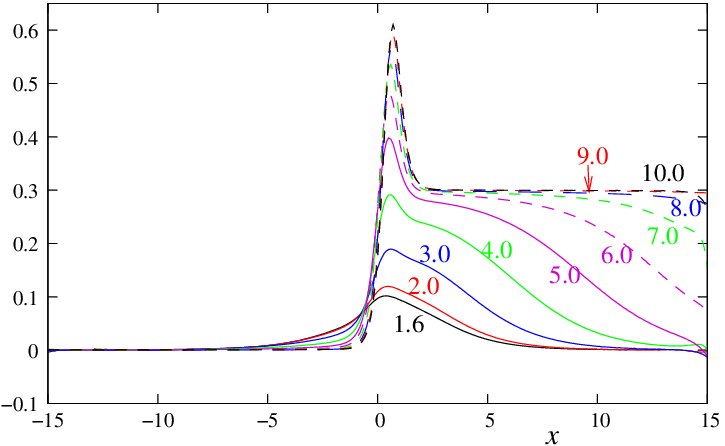}
\caption{The kinetic and response components of the correlation potential $v\ckin + v\cresp$ for our model with $c=0.6$; the values of interatomic separation $R$ are indicated.}
\label{fig:vkin+respa}
\end{figure}

\begin{figure}[h]
 \centering
 \includegraphics[height=5cm,width=7cm]{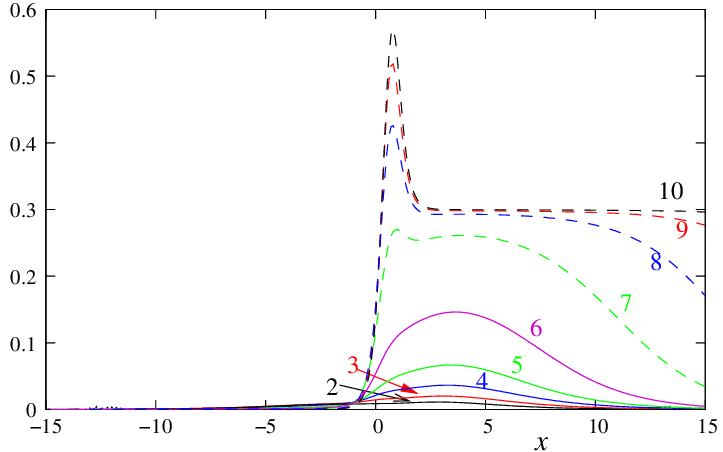}
\caption{As in Fig.~\ref{fig:vkin+respa} but with $c=2.8$}
\label{fig:vkin+respc2.8}
\end{figure}

\begin{figure}[h]
 \centering
 \includegraphics[height=5cm,width=7cm]{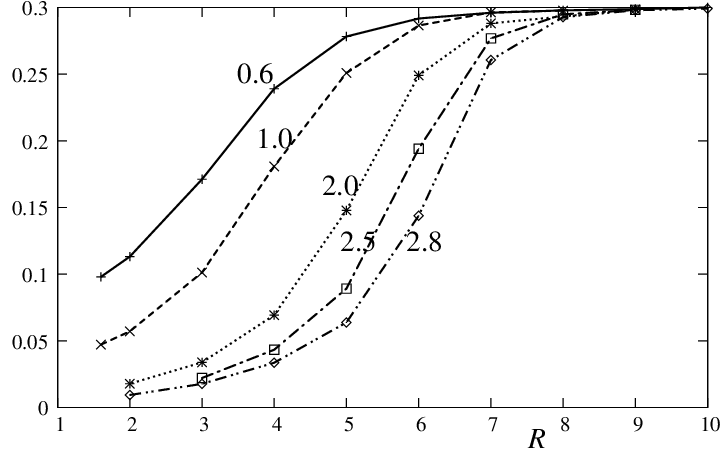}
\caption{The value of $v\ckin + v\resp$ at the location of the H atom in our model, as a function of the internuclear separation $R$, and with $c$-values as indicated.}
\label{fig:vckin+respatH}
\end{figure}

The long-rangedness of the density in soft-Coulomb systems means that
the Hartree and exchange terms decay slower than in the usual Coulomb
case. To clarify the step and peak structures, we plot just the
kinetic plus response term in Figure~\ref{fig:vkin+respa} for our LiH
model of $c=0.6$, and in Figure~\ref{fig:vkin+respc2.8} for $c=2.8$.
The relation between the $R$ at which the step develops and the
avoided crossing discussed above is seen more clearly in these
figures. Finally, in Fig.~\ref{fig:vckin+respatH}, we plot the value
of $v\ckin + v\resp$ at the location of the atom with the larger IP
(the H atom in our model), as a function of internuclear separation
$R$, for various different $c$-values. This graph shows quite clearly
that the development of the step tracks the location and sharpness
of the avoided crossing: the
larger the separation at which the avoided crossing occurs
(i.e. larger $c$-value), the consequently larger $R$ the step is
onset, and that the step develops more sharply, corresponding to the
sharper avoided crossing at larger distances.

%For example, half the asymptotic step is reached at $R$ around 3.0 for c=0.6, $R\sim 4$ for c=1.0, and $R\sim 5...$ for c=2.0.

%\begin{figure}[h]
% \centering
% \includegraphics[height=6cm,width=7cm]{Figures/vkin+respe.eps}
%\caption{The  kinetic and response components of the correlation potential $v\ckin + v\cresp$ for our model but with $c=1.0$}
%\label{fig:vkin+respe}
%\end{figure}

\subsection{The asymptotic separation limit, and the significance of the peak}
\label{sec:asymptotic}
Analytic expressions for the xc potential and its components in our two-electron model can be found in the separated-atom limit, by adopting
the Heitler-London form for the wavefunction:
\ben
\Phi_{HL}(x,x') = \frac{\phi_H(x)\phi_{Li}(x')+\phi_{Li}(x)\phi_H(x')}{\sqrt{2(1+S_{H,Li}^2)}}
\een
where $S_{H,Li}$ is the overlap integral:
\ben
S_{H,Li} = \int \phi_H(x) \phi_{Li}(x)dx
\een
We will focus on the interatomic region, far from either nuclei, in this limit.
To lowest order in the separation $R$, the orbitals $\phi_{Li}(x)$ and $\phi_H(x)$ in this region may be written:
\bea
\nonumber
\phi_H(x)&=&\sqrt{\alpha_{H}}e^{\alpha_{H}(x-\frac{R}{2})}\\
\phi_{Li}(x)&=&\sqrt{\alpha_{Li}}e^{-\alpha_{Li}(x+\frac{R}{2})}
\label{eq:orbitals}
\eea
where $\alpha = \sqrt{2I}$, with $I$ being the first ionization
potential of the atom.
Similar expressions hold for the three-dimensional case, with Coulomb interaction;  the only differences being that instead the orbitals have asymptotic dependence according to
\ben
\phi(x)={\alpha}^{\frac{3}{2}} e^{-\alpha \sqrt{{(x \pm \frac{R}{2})}^2 + y^2 + z^2}}
\een
(where the x-axis is taken to be the bond axis).

%%David please fill in ...above. Maybe even give the equations with a general ``d'' in there, so from the outset it applies to either 1d soft-Coul or 3d Coul.

%For our analysis, we set the ionization
%potentials in Eqs.~(\ref{eq:Hydrogen}) and~(\ref{eq:Lithium}) equal to
%the values used in our numerical model, to reproduce the same
%asymptotic dependence of the density.

It is a simple exercise to construct the first-order density-matrix
and the density using these orbitals.  Substituting
into Eqs.~(\ref{eq:kohn-sham}) and~(\ref{eq:vkin}) yields the
large-separation limit of the KS potential and $v_{\rm kin}(\br)$.  
%We note that the soft-Coulomb orbitals (and their Hartree-xc potentials) are longer-ranged than their 3D Coulomb counterparts, so achieve their asymptotic forms at larger distances.

%The density
%obtained in this way can then be substituted into
%eq.~(\ref{eq:kohn-sham}) to obtain an approximate asymptotic analytic
%expression for the KS potential. Similarly, the first order
%density matrix can be substituted into eq.~(\ref{eq:vkin}) to obtain
%an approximate analytic expression soley for $V_{Kin}(r)$.

%The
%expressions obtained in this way become exact at all separations for
%delta-function potentials and correct in the asymptotic limit for the
%soft-Coulomb potentials we are interested in analyzing.

%%David - for delta-fn potential, i don't think that's right above that it is exact at all separations, bceause of Hxc effects, agree? Yes, I agree!

%In the large separation limit for a diatomic system,
%the Hartree
%repulsion is expected to be relatively small in the inter-atomic
%region.
In the limit of large interatomic separation, the Hartree potential vanishes as the inverse distance from the nuclei in the inter-atomic region.
Also, in this limit, the second order density matrix
factorizes into a product of densities and it follows from
Eq.~(\ref{eq:vxchole1}) that $v\xc^{\rm hole}(r)$ also falls off as the inverse distance from the nuclei
in the interatomic region.
The KS potential is then dominated by contributions from $v\kin(r)$ and
$v\resp(r)$. Explicitly, in one dimension this is given by:
\bea
\nonumber
v\s &=& \frac{1}{2} \frac{|\phi_H'|^2 +
|\phi_{Li}'|^2
+2\sqrt{\epsilon}(\phi_H')(\phi_{Li}')}{|\phi_H|^2
+ 
|\phi_{Li}|^2 + 2\sqrt{\epsilon}\phi_H \phi_{Li}}\\
\nonumber
&+&\frac{1}{2}\frac{\phi_H\phi_H''+\phi_{Li}\phi_{Li}''+\sqrt{\epsilon}(\phi_{Li}\phi_H''+\phi_{H}\phi_{Li}'')}{|\phi_H|^2 +
|\phi_{Li}|^2 + 2\sqrt{\epsilon}\phi_H \phi_{Li}}  
\\
&-&\frac{1}{4}
\frac{(\phi_H\phi_H'+\phi_{Li}\phi_{Li}'+\sqrt{\epsilon}(\phi_{Li}\phi_H'+\phi_{H}\phi_{Li}')^2}{(|\phi_H|^2
+ |\phi_{Li}|^2 + 2\sqrt{\epsilon}\phi_H \phi_{Li})^2} -I_{Li}
\eea
In the above expression $\phi'$ and $\phi''$ denote the first and second spatial derivatives of the orbital, and  $\epsilon$ is the square of the overlap
integral at interatomic separation $R$:
\ben \epsilon = \frac{\alpha_H \alpha_{Li}}{(\alpha_H -
\alpha_{Li})^2}(e^{-R \alpha_{Li}} - e^{-R \alpha_{H}})^2 \een

%In the 3-d case, the above expression for $v\s$ is the same,
%provided the derivatives are replaced by gradients (and dot products
%etc. where needed) and $\epsilon$ is replaced by the square of the
%3-dimensional overlap integral. (See handwritten notes for the 3-d
%overlap integral...it's very messy, but we do have an explicit
%analytic form if needed).

\begin{figure}[ht]
 \centering
 \includegraphics[height=8.5cm,width=7cm]{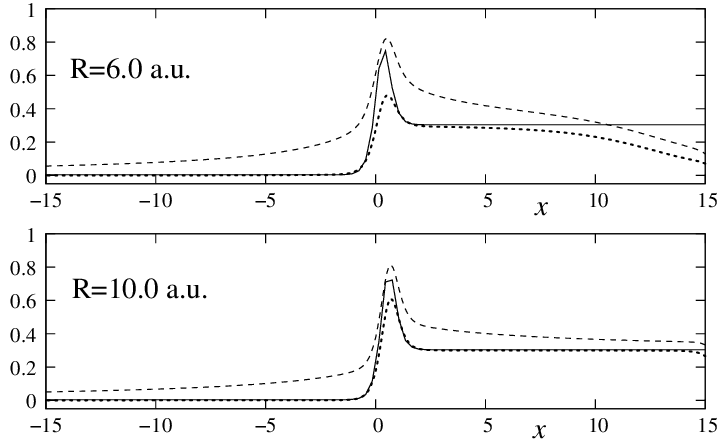}
\caption{Asymptotic expression for $v\Hxc$ (solid curve) in the 
inter-atomic region compared to $v\Hxc$ in our model (dashed curve),and $v\kin + v\resp$ (dotted line) in our model at separations indicated.}
\label{fig:asymp}
\end{figure}

In Fig.(\ref{fig:asymp}), the asymptotic expression for $v\Hxc (= v\s - v\ext)$ using the orbitals of Eqs.~\ref{eq:orbitals}
is plotted for comparison with the
$v\ckin(r)+v\resp(r)$ component of the numerical solution using the
soft-Coulomb potentials.
(As noted earlier, the soft-Coulomb orbitals are longer-ranged than their 3D Coulomb counterparts, so $v\Hxc$ achieves its asymptotic form only at larger distances.)
 We see that the step
reaches its asymptotic limit more quickly than the peak. For
instance, at $R=10.0$ a.u. the peak for the numerical solution is
somewhat smaller than that of the analytic expression, although the
step has already reached its asymptotic value of $3.0$H.
%\begin{figure}[ht]
% \centering
% \includegraphics[height=5cm,width=7cm]{fig6.eps}
%\caption{Analytic expression for the KS potential (solid curve) and
%the $v_{Kin}(r)+v\resp(r)$ components of the numerical solution
%(dashed curve) at a) R=4.0 a.u. b) R=8.0} \label{fig:vanvnum}
%\end{figure}

We next derive asymptotic
expressions for the location and magnitude of the peak and step
structures as functions of the internuclear separation. Defining the
location of the peak from the condition
$\frac{d}{dx}v\ckin\vert_{x_{peak}} = 0$, we obtain \ben
x_{\rm peak}=\frac{R}{2}\frac{(1-\sqrt{\frac{I_{Li}}{I_H}})}{(1+\sqrt{\frac{I_{Li}}{I_H}})}+\frac{1}{\sqrt{32}}\frac{ln\frac{I_{Li}}{I_H}}{\sqrt{I_{Li}}+\sqrt{I_H}}\label{eq:peakloc}
\een
where $I_{Li}$ and $I_H$ are respectively the ionization potentials
of Li and H. 
 For the 3D case the second term on the right
is modified to be:
\ben
\frac{3}{\sqrt{32}}\frac{ln\frac{I_{Li}}{I_H}}{\sqrt{I_{Li}}+\sqrt{I_H}}
\label{eq:peaklocl3d}
\een
Defining the location of the step by its inflection
point, i.e. from the condition $\frac{d^2}{dx^2}v\resp=0$, one
obtains the same result, i.e. \ben
x_{\rm step} = x_{\rm peak} \;.
\label{eq:steploc}
\een
Therefore, in
the asymptotic limit, our two-electron model shows that  the location of
the peak and step coincide. 
The second term in Eq.~(\ref{eq:peakloc}) is negative, but
in general small compared to the first term for large inter-atomic
separation $R$. Therefore, the peak and step structures are located
closer to the hydrogen atom; more generally, closer to the more
electronegative atom of a diatomic molecule.
On the other hand, the minimum of the density:
\ben
x_{\rm min(n)}=\frac{R}{2}\frac{(1-\sqrt{\frac{I_{Li}}{I_H}})}{(1+\sqrt{\frac{I_{Li}}{I_H}})}+\frac{1}{\sqrt{32}}\frac{ln\frac{{I_{Li}}^2}{{I_H}^2}}{\sqrt{I_{Li}}+\sqrt{I_H}}
\label{eq:densmin}
\een
lies closer to Li than the
peak/step location, but still on the side of the bond mid-point closer
to H: 
The first term of Eq.\ref{eq:densmin} is identical
to Eq.~(\ref{eq:peakloc}), while the second term contains the
logarithm of the ratio $\frac{{I_{Li}}^2}{{I_H}^2}$ instead of the
ratio $\frac{I_{Li}}{I_H}$, which is smaller than one. 

Our simple two-electron model thus explains the earlier observations 
in real molecules~\cite{GB96}: In the general many-electron
hetero-atomic case, given that the peak and step structures arise from the
delocalized HOMO, our analysis can predict their positions. 
The location of the step was seen to coincide
with the peak in the true LiH molecule, with both lying closer to the H atom, 
at least for the largest interatomic distances that those calculations were able 
to perform. 
For the homo-atomic case, our results 
(Eqs.~\ref{eq:peakloc} and~\ref{eq:densmin}) predict that 
$x_{\rm peak}=x_{\rm dens,min} = 0$ and so the minimum of the density and peak
location coincide at the bond midpoint; also borne out by the examples
in the literature.

%The magnitude of the peak in the correlation potential of the
%numerical model is plotted in fig.(\ref{fig:peak}) for comparison
%with the peak in the KS potential of our analytic expression. 
%The
%two curves will eventually coincide in the infinite separation limit
%to the value 0.7672 a.u. This value contains contributions from both
%the peak in $v_{Kin}(r)$ and the step in $v\resp(r)$. 
We next turn to the magnitudes of the structures. 
Using the
density matrix constructed from the orbitals in
Eqs.~\ref{eq:orbitals}, one can show that the
magnitude of the peak structure in $v\kin(r)$, in the limit
that the overlap integral vanishes, is given by the expression: 
\ben
v\ckin^{max} = \frac{1}{4}\left(\sqrt{I_H} + \sqrt{I_{Li}}\right)^2
\een 
For our two-electron model of LiH, this gives a value of 0.616
a.u. Adding the value of the step in $v\resp$ at its inflection
point ($\Delta I/2 = 0.15$au), gives .7672 a.u., which is indeed what the peak of our numerical solution asymptotes to. 
For
the homo-atomic case, the above expression gives a value of
$V_{\rm peak}=0.5$au, agreeing with the results of Refs.~\cite{GB96}
and~\cite{BBS89} for the true homo-atomic two electron system H$_2$.
However, in Ref.~\cite{GB96}, the magnitude of the peak for
{\it true} LiH, was significantly smaller than this prediction. This
discrepancy is due to the effect of the localized core electrons in the Li 1s
shell, which lead to a dramatic decrease in the magnitude of the
gradient of the conditional probability amplitude
eq.~(\ref{eq:condit}) in the inter-atomic region, and hence by
eq.(~\ref{eq:vkin}), a decrease in the magnitude of the peak.

%\begin{figure}[ht]
% \centering
% \includegraphics[height=9cm,width=7cm]{15.eps}
%\caption{Peak magnitude in $V_{c}$ for the numerical model as compared with our analytic expression for the KS potential}
%\label{fig:peak}
%\end{figure}
As discussed in Sec.~\ref{sec:decomposition}, the peak emerges out
of analyzing the change in the conditional probability. We now give
a different argument for why the peak must be there, even though it has negligible
effect on the ground-state energetics. 
The peak occurs when one takes the "non-bonding" orbital as the KS orbital:  
\ben
\phi = \sqrt{(\rho_H + \rho_{Li})/2}. 
\een
(Here $\rho_H$ is the atomic density of the H atom and $\rho_{Li}$ that of the Li atom, i.e. the squares of the orbitals in Eq.~\ref{eq:orbitals}). 
This is the exact doubly-occupied KS orbital,  since twice its 
square yields the exact density in the limit of infinite separation, $\rho = 2\vert\phi\vert^2 = \rho_H +\rho_{Li}$.

\begin{figure}[ht]
 \centering
 \includegraphics[height=5cm,width=7cm]{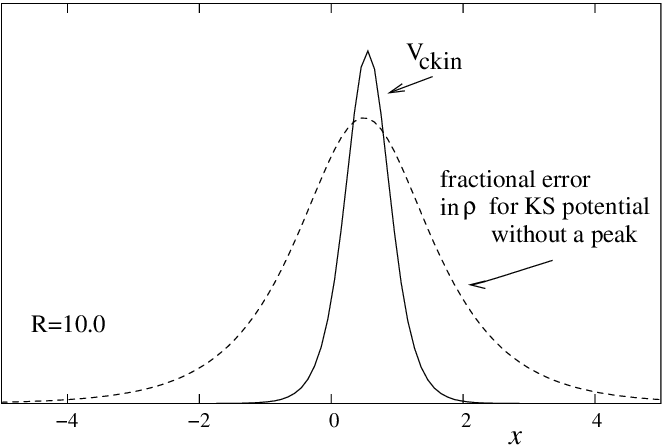}
\caption{The peak in $v\ckin(x)$ and that in the fractional density-error of the KS orbital solution to the KS potential with the peak taken out, $\sqrt{\rho_H\rho_{Li}}/(\rho_H+\rho_{Li})$, (see text) have the same location.}
\label{fig:peaks}
\end{figure}

If one instead takes the "bonding orbital": 
\ben
\phi_{\rm bond}= (\sqrt{\rho_H/2} + \sqrt{\rho_{Li}/2}), 
\een
and finds the KS
potential corresponding to this, there is {\it no} peak
structure (but there is still the step). 
That is, if one asks what is the KS orbital for the KS potential with the
peak structure sliced out, the KS orbital would instead be $\phi_{\rm bond}$.
Now the density corresponding to
$\phi_{\rm bond}$ is
\ben
\rho_{\rm bond} = 2\vert \phi_{\rm bond}\vert^2 = \rho_H + \rho_{Li} + 2\sqrt{\rho_H\rho_{Li}}
\een
i.e.  is equal to the sum of the atomic densities
{\it plus} a term $ 2\sqrt{\rho_H\rho_{Li}}$. This term is indeed very small, 
but taken as a fraction of the total density, 
$\sqrt{\rho_H\rho_{Li}}/(\rho_H+\rho_{Li})$, displays a peak at the exact
same location as the peak in the {\it exact} KS potential, Eq.~\ref{eq:peakloc} (Figure~\ref{fig:peaks}). 
 The shape of the peak is 
different but its maximum coincides in the limit of infinite separation.
This suggests an interpretation of the peak in $v\ckin$ (in the exact KS potential), as 
a barrier that pushes back to the atomic regions the extraneous density $2\sqrt{\rho_H\rho_{Li}}$
that would be in the bonding region if the peak was absent. 
Since the KS system by definition must get the density correct the peak
must be there.

The interpretation here is closely related to the analysis of
Ref.~\cite{GB97} of homo-atomic molecules, where it was shown that the
kinetic energy density for the exact KS orbital develops a well in the
bond mid-point region, that must be compensated by a peak in the KS
potential in order to keep the constant value of the KS orbital
energy. An LCAO approximation to the orbital (analogous to
$\phi_{\rm bond}$ above) does not display the well.

\section{Discussion and Implications}
\label{sec:implications}
Using a simple one-dimensional model of a two-electron heteroatomic
molecule we studied features of the exact KS potential that arise for
real 3D heteroatomic molecules.  In particular we examined the
characteristic step and peak structure in the internuclear region,
that develop as the molecule dissociates.  These unusual features are
a peculiarity of the non-interacting KS description: on the one hand,
as a molecule dissociates, the interaction between the electrons on
one atom and those on the other vanishes, so why, fundamentally, do
such stark structures appear in the KS potential?  The answer
ultimately lies in the single-Slater-determinant description in the KS
system: although this is indeed how the exact KS system describes the
state, it is far from the true wavefunction which needs, even
qualitatively, two Slater determinants.  In the two-electron model, the KS system consists of a doubly-occupied spatial orbital, blatently far from the true two-orbital interacting system. 
Mathematically, the structures can be
understood by considering the response and kinetic components of the
correlation potential, as explained in earlier works and in
Section~\ref{sec:decomposition} of the present paper.  Physically, a KS potential that lacks the step leads to dissociation into fractional charges; a KS potential
that lacks the peak leads to a KS orbital that yields an incorrect (albeit exponentially small) density in the internuclear region. The former point is well-recognized in the literature, while the latter point elaborates on an earlier interpretation~\cite{GB97} (Sec.~\ref{sec:asymptotic}). 
 
Due to the simplicity of our two-electron model, we are able to
investigate in much more detail than in the earlier literature, the
development of these structures and their asymptotic properties.
Several of these features carry over to the true many-electron 3D
case, since they arise from the HOMO orbital.  We showed that the step
begins to develop at the internuclear separation where the avoided
crossing in the ground and lowest charge-transfer state is approached,
and explained why. We gave an exact formula for the location of the
step and peak, in the limit of large separation, finding that the two
structures are located at the same place, and closer to the atom with
the larger IP, consistent with the few calculations done on real
molecules in the literature.

Being in a region of very low electron-density, these features, in
themselves, have little energetic consequences for the ground states
of these systems.  However they have dramatic consequences for
time-dependent processes, excitations, and response. For example, it
has been shown that the related peaks that appear in the interatomic
regions of a hydrogen chain significantly (and correctly) reduce the
polarizability of the chain and that local and semi-local
approximations which lack the peak, consequently significantly
underestimate the (hyper-)polarizability~\cite{KKP04}. As TDDFT begins
to be utilized in molecular transport calculations, we anticipate the
peaks will act as barriers decreasing the current.
%Exact exchange within KLI
%underestimates the peaks, compared to full exact-exchange, and
%consequently the polarizability is not as much reduced as it should
%be.  (*not a correlation effect though, but that's ok*)

The step in the KS potential ultimately imposes a rather complicated
structure on the exact xc kernel of TDDFT~\cite{M05c,MT06}. Because of
the realignment of the atomic HOMO's, the molecular HOMO and LUMO are
symmetric and antisymmetric combinations of the atomic HOMO's, separated in energy merely by
the tunnelling factor, that vanishes as $\exp(-{\rm const.}R)$ as the
molecule dissociates. Therefore three KS determinants become
near-degenerate: the doubly-occupied HOMO, a single-excitation to the
LUMO, and a double-excitation to the LUMO. That is, the step
introduces {\it static correlation} in the KS system that is not
present in the true interacting system. It is the job of the TDDFT xc
kernel to ``undo'' this static correlation, in order to yield good
excitation energies in the true system. This has a dramatic effect on
the structure of the xc kernel for charge-transfer excitations in
molecules composed of open-shell fragments~\cite{M05c,MT06}; in
particular, the double-excitation induces a
strong-frequency-dependence on the kernel.

\begin{figure}[ht]
 \centering
 \includegraphics[height=5cm,width=7cm]{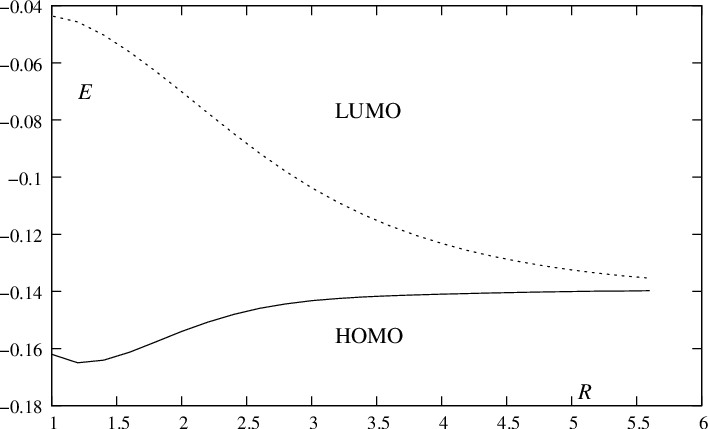}
\caption{LSD KS eigenvalues for LiH become near-degenerate as a function of internuclear separation $R$}
\label{fig:HOMOLUMO}
\end{figure}

Almost all the approximations in use today do not capture the step and
peak structure in the potential. Carefully constructed orbital functionals for the
correlation potential may display these structures, as has been explicitly shown in Ref.~\cite{GB06}.  
Interestingly, static correlation in the KS system
is nonetheless not escaped in the usual (semi-)local
approximations. Delocalized orbitals underlie the fractional charge
problem, and the HOMO and LUMO become near-degenerate as the molecule
dissociates. Figure~\ref{fig:HOMOLUMO} demonstrates this for the LiH
molecule within LSD; a similar merging of the HOMO and LUMO is also
seen in GGA.  
%So if one wanted to build kernels on top of the usual
%GGA's etc, you'd also need strong freq-dep...

%NTM and DGT gratefully acknowledge 
This work is financially supported by the National Science Foundation NSF CHE-0547913 (NTM \& DGT), NSF CHE-07-19291 (TJM), a Research Corporation Cottrell Scholar Award (NTM), and the Hunter Gender Equity Project (NTM \& DGT).

%\section{References}

\end{document}